\documentstyle[11pt]{article}

\begin{document}
\topmargin 0pt
\oddsidemargin 5mm
\setcounter{page}{1}
\begin{titlepage}
\hfill Preprint YERPHI-1540(14)-99

\vspace{0.5cm}
\begin{center}

{\bf $P$-Parity Violating Bound States of Particles with Anomalous Magnetic Moment\footnote{Short version of this paper has been reported on conference
"Physics-99" (Yerevan State Univ. 13 September 1999)} }\\
\vspace{5mm}
{\large R.A. Alanakyan \\}
\vspace{2mm}
{(C) All Rights Reserved 1999\\}

\vspace{5mm}
{\em Theoretical Physics Department,
Yerevan Physics Institute,
Alikhanian Brothers St.2,
Yerevan 375036, Armenia\\}
 {E-mail: alanak@lx2.yerphi.am\\}

\newpage
\end{center}
\vspace{3mm}
\centerline{{\bf{Abstract}}}

We consider bound states of fermions with an anomalous magnetic moments(neutrinos,neutrons)
in radial electric and magnetic  field of monopole.In case of the
radial magnetic field the interaction $\vec{\Sigma}\vec{H}$ violates $P$-parity and for this
reason we must use the method of \cite{RA7}(hep-ph/9901248 ) where both components of spinor considererd
 as a linear combination of spheric spinors with different $P$-parity.Also we apply
pseudoscalar-like interaction (2) obtained in \cite{DF}
to monopole case and add it in Dirac equation.We obtain the system of differential
equations for radial functions which define energy levels of fermions
with  anomalous magnetic moments in the presence of monopole.

\vspace{3mm}
\vfill
\centerline{{\bf{Yerevan Physics Institute}}}
\centerline{{\bf{Yerevan 1999}}}

\end{titlepage}

As known, the dynamics of the neutral  fermions with anomalous magnetic
moments are described  by Dirac equation with nonminiaml coupilngs
of neutral fermions with electromagnetic field \cite{LL4},\cite{AB}:
\begin{equation}
\label{A18}
(\hat{k}-m_n+\mu_n (\vec{\Sigma}\vec{B}-i\vec{\alpha}\vec{E})
+iq(\vec{E}\vec{B})\gamma_5 )\psi(k)=0,
\end{equation}
where  $\mu_n $-is anomalous magnetic moment,
$\frac{1}{2}\vec{\Sigma}$ is spin operator,and defined by formula (21,21) of \cite{LL4}
operator $\vec{\alpha}$ is defined by formula (21,20) of \cite{LL4}.

In this equation we include also term obtained in ref. \cite{DF}:
\begin{equation}
\label{A4}
L=iq(\vec{E}\vec{H})\bar{\psi} \gamma_5 \psi
\end{equation}
which appears e.g. in electroweak models at one-loop level in theories with $P$-parity
violation(for electroweak thories and $P$-parity violation see e.g. \cite{O} and
references therein).

In this article  we consider bound states
 of particles with anomalous magnetic moments  and with interaction (2) in the presence of
 monopole (see \cite{P1} \cite{H1}  references in \cite{R}).
 We obtained  generalization of equations (13),(14) \cite{RA1} where has been considered
 bound states of particles with anomalous magnetic moments in arbitrary radial electric field.

In \cite{RA1} has been considered joint influence of the static radial electric field
 and magnetic fild.Although in this paper $B$ is not radial as has been shown that
in some condition term $\vec{\Sigma}\vec{B}$ is radial.However in contrast to monopole case
in \cite{RA1}  term  $\vec{\Sigma}\vec{B}$ is $P$-parity  conserved.

The magnetic field (but not magnetic field of monopole) has been presented in equations
(13),(14) of ref. \cite{RA1} besides radial electric field.
During derivation of equations (12)-(15) below which defines
energy levels of the fermions with anomalous magnetic moment in the presence of
monopole the method  of the \cite{RA7}
has been used
(because in case of radial  magnetic field the term $\vec{\Sigma} \vec{B}\sim \vec{\Sigma} \vec{r}$
violates $P$-parity) for angular variables separation, in accordance with
this method
it is necessary to present components of spinors as linear combination  of spheric spinors
$\Omega_{jlM}(\vec{n}),\Omega_{jl'M}(\vec{n})$ which have different $P$-parity.
We also confirm the result
of \cite{RA1} where has been stressed that in case of Coulomb  electric field take place the
fall down on the center takes place.

As known(see (see e.g.  references in \cite{R})) exist nontrivial
solutions of to Yang-Mills theories which have in general both
electric and magnetic fields:
\begin{equation}
\label{A5}
\vec{B}=\vec{n}B(r)
\end{equation}
\begin{equation}
\label{A5}
\vec{E}= \vec{n}E(r)
\end{equation}
where$\vec{n}=\frac{\vec{r}}{r}$.At large distances from the core
of the monopole we have:
\begin{equation}
\label{A5}
\vec{E}=\frac{e \vec{r}}{r^3}
\end{equation}
\begin{equation}
\label{A5}
\vec{B}=\frac{g \vec{r}}{r^3}
\end{equation}
It must be noted that between between electric and magnetic charges
there exist a relation like that between electric and magnetic
charges in case of Dirac monopole \cite{D1} :
\begin{equation}
\label{A5}
eg=\frac{1}{2}n, \quad n=0,\pm 1,\pm 2,\pm 3,...
\end{equation}

Thus below we consider bound states of neutral fermions with
anomalous magnetic moment in both radial magnetic and electric fields
created by these objects.

Also it is of interest to note that in case of monopole  $\vec{E}\vec{B}=E(r)B(r)$ and
thus pseudoscalar interaction (2) is spherically symmetric (depends only on $r$)
  as well as the interaction
connected with anomalous magnetic moment and separation of
angular variable  is possible as  seen below.At large
distances:
\begin{equation}
\vec{E}\vec{B}=E(r)B(r)=eg \frac{1}{r^4}
\end{equation}
and thus we have  radial interaction (2) which is attractive at appropriate sign
of $n$ and lead to fall down on the center.The non-locality of
$L=iq(\vec{E}\vec{B}) \bar{\psi} \gamma_5 \psi$ vertex and finite size of
monopole prevent this fall down.Besides, the term:
\begin{equation}
\label{A5}
q^2(\vec{E}\vec{B})^2=q^2 \frac{e^2g^2}{r^8}
\end{equation}
in effective potential is always repulsive and dominates
at small distances.

It must be stressed that $\vec{E}\vec{B}$ depends only on $r$
because $\vec{B}$ is radial.
Also radial is the interaction of particles with anomalous magnetic
moment with electric and magnetic fields of monopole in Dirac equation
for particles with anomalous magnetic moment :
\begin{equation}
\label{A18}
(\hat{k}-M(r)+ \mu (g\vec{\Sigma}\vec{n}B(r)-ie\vec{\alpha}\vec{n}E(r))+iqE(r)B(r)\gamma_5 )\psi(k)=0,
\end{equation}
Here $M(r)=c \phi(r)$ ($\phi(r)$- is Higgs field which give mass to the fermion, $m=M(\infty)$-is the
observed mass of fermion).

Here $P$-violation presented
(because  interaction (2) is effectively pseudoscalar,
$\vec{\Sigma}\vec{n}B(r)$ also $P$-odd )and we find as  in [10] the solution as
 linear combinations of spheric spinors $\Omega_{jlM}(\vec{n}),\Omega_{jl'M}(\vec{n})$
which have different $P$-parity:
\begin{eqnarray}
\label{A20}
&&\psi^T=
(\phi,\chi),\nonumber\\&&
\phi=f_1(r)\Omega_{jlM}(\vec{n})+
(-1)^{\frac{1+l-l'}{2}}
f_2(r)\Omega_{jl'M}(\vec{n}),\nonumber\\&&
\chi=g_1(r)\Omega_{jlM}(\vec{n})+
(-1)^{\frac{1+l-l'}{2}}
g_2(r)\Omega_{jl'M}(\vec{n}))
\end{eqnarray}
After separation of angular variables we obtain the following set of equations for
radial functions:
\begin{equation}
\label{A27}
f_1'(r)+\frac{1+\kappa}{r}f_1(r)+ \mu E(r)f_1(r)=(\epsilon+M(r))g_1(r)
-i \mu B(r)g_2(r)+a_P(r)f_2(r)=0
\end{equation}
\begin{equation}
\label{A26}
f_2'(r)+\frac{1-\kappa}{r}f_2(r)+ \mu E(r)f_2(r)=-(\epsilon+M(r))g_2(r)
-i \mu B(r)g_1(r)-a_Pf_1(r)=0
\end{equation}
\begin{equation}
\label{A25}
g_1'(r)+\frac{1-\kappa}{r}g_1(r)- \mu E(r)g_1(r)=-(\epsilon-M(r))f_1(r)+i \mu B(r)f_2(r)
+a_Pg_2(r)=0
\end{equation}
\begin{equation}
\label{A24}
g_2'(r)+\frac{1+\kappa}{r}g_2(r)- \mu E(r)g_2(r)=(\epsilon-M(r))f_2(r)+i \mu B(r)f_1(r)-a_Pg_1(r)=0
\end{equation}
where
\begin{equation}
\label{A23}
\kappa=l(l+1)-j(j+1)-\frac{1}{4},
\end{equation}
\begin{equation}
\label{A30}
a_P=iqE(r)B(r),
\end{equation}
At  large distances $a_P= \frac{qeg}{2r^4}$.

In pure electric field case i.e. at $B=0, q=0$ $P$-parity is conserved
 and we obtain two decoupled system of equations
(13),(14) of the [9] in which of course the magnetic field is also zero.

If only electric field is presented we obtain for radial functions the following equations:
\begin{equation}
\label{A27}
(\vec{p}^2+m^2-\epsilon^2+\mu^2E^2+4\mu\pi\rho \pm \frac{2\mu E(r)(1+\kappa)}{r})R_{1,2}(r)=0.
\end{equation}
where $\vec{p}^2=-\frac{1}{r^2}\frac{d}{dr}r^2\frac{d}{dr}+\frac{l(l+1)}{r^2}$
(we find the solution as $\phi=R_1(r)\Omega_{jlM}(\vec{n}),\chi=R_2(r)\Omega_{jl'M}(\vec{n})$ ).
During derivation of this formulas we take into account that $div \vec{E}=4 \pi \rho$.

In \cite{RA1} it has been stressed (page 3 after formulas (13),(14)) that for particles
with anomalous magnetic moment we have fall down on the center in case of Coulomb electric
field which prevented by cut off of the potential, i.e. by taking into account charge
distribution inside neutron.

Indeed, from equations (18)-(19)(as well as from equations (13)(14) of the ref.\cite{RA1})
 it is seen that in case of Coulomb attraction ($E(r)=\frac{Ze}{r}$)
we have fall down on the center due to the term
$\mu\frac{E(r)}{r}(1+\kappa)=\frac{\mu eZ(1+\kappa)}{r^3}$in effective potential.
It must be noted, however, that the
term $\mu^2E^2=\frac{Z^2\mu^2e^2}{r^4}$ in potential is always repulsive and at
small $r$ prevents fall down on the center.
In case of pointlike charge term $2\pi\rho=2\pi\mu Ze \delta(\vec{r})$
may be considered as perturbation.

In case of Coulomb potential we can calculate by using quasiclassical methods energy levels
which are defines by the equation \cite{LL3}:

\begin{equation}
\label{A27}
\int\sqrt{2m(E-V(r)}dr=n
\end{equation}

where
\begin{equation}
\label{A27}
V(r)=\frac{(l+\frac{1}{2})^2}{2mr^2}+\frac{\mu E(1+\kappa)}{r}+\frac{1}{4}\mu^2E^2
\end{equation}

\begin{equation}
\label{A27}
2mE=\epsilon^2-m^2
\end{equation}

{\bf Appendix}

Below we present system of equation for radial functions (19)-(22) of ref. [10]
which obtained after angular variables separation in electroweak Dirac equation:
\begin{equation}
\label{A27}
(f_1'(r)+\frac{1+\kappa}{r}f_1(r))-(E+M(r)-V(r))g_1(r)-V_+(r)f_2(r)=0
\end{equation}
\begin{equation}
\label{A26}
(f_2'(r)+\frac{1-\kappa}{r}f_2(r))+(E+M(r)-V(r))g_2(r)+V_+(r)f_1(r)=0
\end{equation}
\begin{equation}
\label{A25}
(g_1'(r)+\frac{1-\kappa}{r}g_1(r))+(E-M(r)-V(r))f_1(r)+V_-(r)g_2(r)=0
\end{equation}
\begin{equation}
\label{A25}
(g_2'(r)+\frac{1+\kappa}{r}g_2(r))-(E-M(r)-V(r))f_2(r)-V_-(r)g_1(r)=0
\end{equation}
where:
\begin{equation}
\label{A28}
M(r)=m-a_SV_S(r)
\end{equation}
\begin{equation}
\label{A29}
V(r)=eg_VZ_0(r)+eQA_0(r),
\end{equation}
\begin{equation}
\label{A30}
V_{\pm}(r)=eg_AZ_0(r)\pm a_PV_P(r),
\end{equation}

The author express his sincere gratitude to E.B.Prokhorenko and Zh.K.Manucharyan
for helpful discussions.

\end{document}